# On a Morphology of Contact Scenario Space

***Joseph Voros***

Senior Lecturer in Strategic Foresight
Faculty of Business and Law
Swinburne University of Technology
John Street, Hawthorn, VIC 3122, Australia
jvoros@swin.edu.au

## Abstract

The purpose of this paper is to explore the possibility space of scenarios of 'contact'—the discovery of extra-terrestrial life, whether intelligent or not—using a morphological approach utilising seven principal parameters, chosen both for their descriptiveness of the scope of possibilities, as well as for their relevance in examining potential societal impacts arising from different scenarios, including, for example, the proximity of the discovery. Several classes of contact scenario are examined, and existing approaches to the search for extra-terrestrial life and intelligence are situated within the range of possible search strategies and targets illuminated by this particular choice of parameters, as are some examples of contact scenarios from popular culture. The resulting possibility space can also suggest new search strategies and potential targets, one of which is highlighted—that of 'galaxy-scale macro-engineering'. It is hypothesised that an example of this might already be known to us, namely the intriguing galaxy 'Hoag's Object' (PGC54559), and some specific empirical observations are suggested in order to test this hypothesis. Some possible extensions to the parameters used, as well as some preliminary observations about modelling the range and extent of human societal *responses* to contact, are also made.

## Highlights:

- Applies morphological methods to model the possibility space of discovering extra-terrestrial life
- Develops a 7-parameter schema 'LSEARCH' for examining possible detection scenarios
- Classifies selected historical and proposed searches, and scenarios drawn from popular culture
- Proposes a novel target for closer empirical observation: Hoag's Object (PGC54559)
- Discusses and outlines possible extensions to the present schema for further future work

## 1. Introduction, Motivation and Background

This paper continues a train of thought outlined in an earlier paper a decade ago, where the idea was raised of conducting a detailed analysis of the parametric space of the possibilities for the discovery or detection of extra-terrestrial life—whether intelligent or not—in order to examine and prepare for the implications of such an event [92]. The method employed is based upon the 'morphological approach' to problem definition and research developed by the legendary astrophysicist Fritz Zwicky during the first half of the 20th century [102], as is described extensively elsewhere in this Special Issue. In brief, since the morphological method is founded upon the systematic enumeration and examination of all conceivable possibilities within a 'possibility space', it is very well-suited as a basis for thinking systematically about the myriad possibilities inherent in the search for extra-terrestrial life or intelligence. A detailed introduction to morphological methods in the context of Futures Studies and thinking about the future possibilities and evolution of social systems, was given in [93], although the reader will not be assumed to be familiar with that work; instead, aspects of the discussion there will be briefly re-iterated here where they are pertinent for our present purposes. In addition, as detailed discussions of the morphological approach can also be found elsewhere in this Issue, we will here mainly focus only on any notable differences from the usual approaches.

Historically, the modern search for extra-terrestrial intelligence (SETI) was conceived, designed and executed from the technological base which our civilisation possessed in the mid-to-latter part of the 20th century CE. It is usually taken as dating from the 1959 "landmark paper" [see, e.g., 26] of physicists Giuseppe Cocconi and Philip Morrison [16] which proposed searching for interstellar (radio) communications, and was a product of our knowledge and understanding of the possibilities of life and intelligence from that era [for reviews of SETI, see, e.g., 53, 80, 82]. Thus, the initial assumption-base and focus of SETI was understandably on biological beings with radio telescopes who might be broadcasting deliberately, or whose domestic electromagnetic radiation broadcasts might be 'leaking' into interstellar space, such as are our own.

Much has occurred in the intervening time, including the arising of the related field of 'astrobiology'—the broad multi-disciplinary study of the possibilities for life (not necessarily intelligent) in the universe [see, e.g., 13, 29, 52, 57, 81], of which SETI could reasonably be considered a specialised sub-activity that is specifically concerned with looking for signs of *intelligence*. There has in recent years been a growing feeling in the SETI community that the search parameters initially laid out historically since the inception of SETI might need widening beyond the default view of searching for forms of electromagnetic radiation possessing certain characteristics emanating from the vicinity of certain types of stars (see, e.g., Davies & Wagner [23], Shostak [77], Ćirković [14], Dick [27], and many references found therein, among numerous others). More recent work in SETI has even argued that *biologically*-based intelligence itself may be a relatively short-lived stage of development for truly long-lived intelligent species [25]. The concept of 'The Singularity' [6, 34, 49, 91]—well-known to futurists since the early 1990s, which posits accelerating bio-info-nano-technological progress leading to human development advancing to a 'post-human' stage within several decades—has been extended to the idea of *post-biological* intelligence in the universe more generally [27, 77]. Thus, it is suggested, conventional or "orthodox" SETI [5], in looking for radio signals from 'suitable' stars that may host 'suitable' planets, may in fact be hunting what might be a very short-lived prey [77]. Hence, it is argued, the SETI enterprise may need complementing with newer concepts and search strategies to widen the possibility of a successful detection. Indeed, SETI pioneer Frank Drake himself has said that the thinking of *futurists* may be needed to help expand our conceptions of what to search for, and where to search, for truly long-lived civilisations [30].

The approaches currently in use in the search for extra-terrestrial life (SETL)—i.e., not necessarily intelligent and so part of the broader astrobiological enterprise, rather than of SETI *per se*—are generally aimed at detecting traces of life either in our own Solar System (e.g., Mars, or the satellites of Jupiter and Saturn) or in the atmospheres of the ever-increasing number of known extra-solar planets ('exoplanets'). SETI scientist Seth Shostak has suggested [76] that these three types of search—the two SETL approaches for what he calls "stupid life", and SETI for *intelligence*—are each likely to yield a positive result within a couple of decades. If so, we had better start preparing ourselves for the consequences of such a detection—what is generically referred to as 'contact' [41, 45]—especially if it turns out to involve intelligent entities [84]. This paper is intended to be a contribution to that process.

The purpose of this paper, then, is to examine the parameter space of scenarios of 'contact'—the discovery or detection of extra-terrestrial life, whether intelligent or not—so as to expand our thinking around the possibilities that may exist for this event as well as forming a basis to consider the implications (broadly conceived) which these possibilities might hold for our civilisation. In the following section (for the sake of completeness and the unity of this paper), some of the properties of the morphological approach as they are relevant to the discussion here will be briefly sketched. Following this, a number of parameters which may be used to characterise the possibility space of contact are discussed and outlined, and an initial set is chosen to demonstrate the method and undertake the preliminary analysis. These are then examined to locate past and current approaches to SETL and SETI within the parametric space, and we also briefly consider contact scenarios as they have been depicted in popular culture. The use of a morphological perspective is able to bring into view other potential search strategies and new directions or targets to pursue, and a particular configuration sub-space of the parametric possibility space is highlighted in this regard. This configuration sub-space is identified with characterising 'galaxy-scale macro-engineering' [96]. It is hypothesised that the beautiful ring galaxy known as 'Hoag's Object' (PGC54559) might perhaps be an example of such galaxy-scale engineering, and some suggestions for more detailed observations of this object for any signs of engineering activities are given. We very briefly consider the next logical follow-up to the analysis of contact scenario space, namely the range of possible scenarios of the human societal *response* to any actual contact, and we conclude with an invitation to interested others to expand and adapt the preliminary schema shown here in order to deepen the conversation around thinking about the implications for human society that 'contact' would have, as well as to, hopefully, take a closer look at Hoag's Object, 'just in case'…

## 2. Outline of Method

As expounded elsewhere in this Special Issue, the 'morphological approach' was developed by Fritz Zwicky beginning in the 1930s [99-102]. At its heart, the morphological approach attempts to systematically examine the entire range of possible combinations of the various attributes or dimensions of the area/topic of interest. In practice, this means attempting to exhaustively list all of the independent dimensions or attributes that may be used to characterise the situation or focus of interest. These dimension/attributes Zwicky called 'parameters', and there could in principle be any number of parameters, each of which could have any number of discrete 'values', which need not necessarily be numerical—so, in this view, the possible range of parameters and their possible values has no *a priori* limit. In essence, *every* aspect of the focus of inquiry could be considered as 'contingent' rather than 'fixed' and therefore open to consideration as a possible parameter with multiple possible values [99]. This sort of 'contingency thinking' immediately primes the mind to open up to a much wider range of possibilities to consider, and when applied to the basic 'shape' or 'morphology' of the 'space of possibility' can lead to novel combinations and potentially new ideas for exploration [102].

In more precise language: Let there be $n$ parameters which taken together characterise an exhaustive description of the area or focus of inquiry. Each individual parameter $p_j$ in the full set of parameters $p_1 \ldots p_n$ has a positive integer $k_j$ of specific values that thereby define it. Both the parameters and the parameter values should be as independent and mutually-exclusive as possible (what we might intuitively characterise as being 'as "orthogonal" to each other as possible'). These $n$ parameters $p_j$ thereby generate an $n$-dimensional combinatorial 'morphological space' or 'field' that consists of every possible combination of every value of every parameter, sometimes also known as a 'Zwicky box' [65]. If one specific value is chosen for each parameter in the full set, this is known as a 'configuration' within the morphological space/field, which is therefore also sometimes known as a ‘configuration space’. The total numerical value of distinct possible formal configurations so generated can be found from the product of all the values $k_j$, so that, therefore, even a relatively small number of parameters with relatively few values can nonetheless lead to a very large number of combinatorial possibilities. In practice, there are techniques that make the process of examining the large number of possibilities much more tractable than might first appear to be the case given these large numbers [see, e.g., 19, 61, 65]. More details about the formal properties of morphological combinatorial spaces can be found in [68, 69], while a slightly-expanded version of the discussion in this section can be found in [93], upon which it is based.

One useful way to represent the morphological space is as a tabular array, with each column representing one parameter, and with each entry within a column representing one of the parameter values. As each parameter may in general have a different number of values from the others, the columns are not generally of the same length. Most recent work on morphological methods has used this form of representation (e.g., Rhyne [59-61, 63], Coyle [17, 19, 20], Ritchey [65, 67, 68]), and this convention is also used here. (Zwicky usually rendered the parameters as rows of varying lengths set out underneath each other [99, 100]; in this regard, also see Godet [38, 39].)

In the variant of morphological analysis known as ‘Field Anomaly Relaxation’ (FAR) developed by Rhyne [59-62, 64], and subsequently used by Coyle and co-workers [17-22], the different parameters (there called Sectors) are each represented by a unique letter, and a limit is imposed of six or seven parameters/Sectors in total, while the specific actual values taken by each of the parameters/Sectors are called Factors. (Consequently, in FAR the tabular parametric array is known as a ‘Sector/Factor array’.) The rationale for this restricted choice of the number of parameters (Sectors) was explained by Rhyne in several of his publications as being, in essence, to aid the process of seeing each configuration as a whole ‘gestalt’ without overloading the capacity of the mind with too many parameter values (Factors) that could not be held simultaneously in mind. The letters are chosen so that the whole set of parameter/Sector letters produces a ‘pronounceable’ word/acronym, which is used as both a ‘meta-language’ and as a mnemonic. Thus, constructs such as ACTIVES, DELTRIG, ESPARC, GEMCORT, HASIKEP, MISFRUD and so on can be found in the FAR-related references given. Such constructs generate a meta-language because one can speak about the entire set of parameters/Sectors using the constructed ‘word’, the individual parameters/Sectors using the specified letter, or the individual parameter values/Factors using a letter/numerical marker pair; while the presence of the letters in the acronym itself acts as a mnemonic to ensure all parameters/Sectors are remembered and considered during analysis. The order of the letters has no further significance other than to make the given sequence of non-repeated letters more-or-less pronounceable. In what follows, we shall make use of the FAR approach by utilising letter names formulated into what is hoped to be a suitably evocative ‘word’.

Of course, given the manner by which the morphological field is constructed by concatenating a set of quasi-independent parameters intended to exhaustively describe an area of inquiry, not all *formal* combinations of the parameter values will necessarily be mutually ‘consistent’, so some reduction is possible in the number of ‘relevant’ configurations in the configuration space. Ritchey [65, 67] terms this reduction process a ‘cross-consistency assessment’, while Rhyne called it ‘filtering out non-coherent configurations’ [61] and Coyle called it ‘eliminating anomalies’ [19]. This form of reduction can be and usually is carried out *a priori*—i.e., before any further more detailed analysis or examination is undertaken of any particular specific parametric configurations considered *in toto*—by examining and considering all possible pairs of all parameter values across all parameters. In essence, any perceived inconsistencies or anomalies in any pairs of parameter values (or indeed triplets, etc, in later analysis) are therefore to be excluded, and once the field of *formal* possibilities in the total morphological space has been reduced to only those that are considered ‘consistent’, one then has what is termed a ‘solution space’ [65, 67] of possible ‘allowable’ ‘coherent’ configurations. This solution space might also be even further reduced subsequently on a case-by-case basis if specific individual configurations *in toto* do not seem plausible, consistent or coherent on closer examination. For our purposes here, though, we shall forego any pre-emptive *a priori* cross-consistency assessments, for the simple reason that we are essentially investigating uncharted territory where ‘unknown unknown’ relationships may exist. It seems prudent, then, in any analyses of candidate morphologies of contact scenario spaces, to err on the side of non-exclusion *a priori* of any putative parameter-value relationships, and consequently to use the maximal combinatorial set of configurations as the basis for investigation, at least initially. Of course, upon closer examination during subsequent analysis we may find what are perceived as non-coherent inconsistencies in particular specific configurations, which we would then exclude, but we would only do so *post hoc* case-by-case rather than, as it were, 'in bulk' *a priori*.

In the choosing of parameters as a basis for a maximal morphological space, the optimal choices are each as independent as possible from the others. For this reason, we shall attempt in the next section to choose parameters that are, as it were, as 'orthogonal' as possible, in order to maximise the extent of the formal possibility space so generated by the particular choice of parameters thus made.

## 3. Some parameters that may usefully characterise 'contact' scenario space

There have been numerous proposals over the years for parameters that could usefully characterise the possibility space of contact [see, e.g., 1, 24, 42, 45, 79, 83]. More recent work by Vidal, in the broad arena of cosmology and astrobiology [88], has explored and synthesised a significantly-increased number of possible parameters than usually considered in earlier work, culminating in a very comprehensive 26-dimensional impact model [89]. Our goal here, though, is to use a more limited number of parameters which could characterise 'contact scenario space' sufficiently well to allow contemplation of the individual scenarios as a 'gestalt', following Rhyne's approach, rather than as a collection of many parameters concatenated together which a more extensive 'exhaustive' approach might require, and which might as a result cause us to lose sight of some of the subtler insights that such a 'gestalt' perspective might yield. To this end, in this section we shall, as noted, focus our attention on the FAR variant of morphological analysis and, in the particular instance of analysis to be discussed here, seven principal parameters/Sectors will be chosen, each with between 2 and 6 values/Factors. Accordingly, one letter from each of these parameters will be chosen to become the letter designating that Sector, and thus to form part of the mnemonic 'word' characterising the morphological possibility space of contact in this particular instance. Of course, there are many, many ways in which we could have chosen the Sectors to examine—here we are focused on those that seem to the author to have especial utility in characterising the scenario space, both for the exploration of the inherent possibilities that may exist there as well as for the examination of the implications for, and potential responses by, human civilisation.

Therefore, utilising in part the earlier work cited above, we shall consider here *three* main aspects of 'contact': the character and composition of the 'sign' which enables or provides the detection or discovery; the form/class and proximity of the contact itself; and the nature, complexity and Kardashev Type of the 'entity' which is detected or discovered. This last parameter may need some explanation for those not familiar with the Kardashev schema [48], and this is done in more detail below.

### *3.1 The Sign*

First we give thought to the type or composition of the 'sign' which is discovered or detected. This is not quite as straightforward as it might appear at first glance. The very mention of 'contact' usually conjures up an image in people's minds—informed, in part, by the many popular-culture scenarios that have been depicted in literature and film (which we shall visit briefly below)—of an intelligent electromagnetic signal from space, such as was depicted in the 1997 film *Contact* itself. But if one stops to consider for a moment what sort of 'sign' would reveal unambiguous evidence of extra-terrestrial life, then it could just as well be an example or living specimen or even fossilised form of an obviously biological entity, or it might be an artefact, or even a piece of engineering or architecture.

So the *composition* of this 'sign' may indeed be an electromagnetic 'signal', as it was originally conceived in the Cocconi and Morrison paper [16] of conventional SETI, but it might equally well instead be something corporeal or physical (i.e., the entity itself, a fossil, an artefact etc), so we need a way to distinguish these two quite distinct cases. The early proposal of Freeman Dyson [31, 32] made in response to Cocconi and Morrison—to look not for electromagnetic *signals* but, rather, for signs of (large-scale!) alien *technology*—is the archetypal example of a different approach to SETI, consequently known as 'Dysonian' SETI, as opposed to the "orthodox" [5] approach of looking for electromagnetic signals. Likewise, and in this vein, Carrigan [9, 10] for example, has recently written of the closely-related idea of 'interstellar archaeology'—searching for evidence of artefacts of intelligent origin. (The extent to which an example of *solar system-scale* engineering activity could be considered an 'artefact' may well be a question of qualitative distinction rather than simply of degree). For our purposes here, and in order to generate as much orthogonality as possible with the other parameters, we shall choose just two values for the *composition* of the sign: *electromagnetic* (i.e., the sign is itself the electromagnetic signal received); and *corporeal/physical* (i.e., the sign is something tangibly physical, perhaps a manifestation of the entity itself or simply a physical 'remnant' or 'trace' of one, perhaps a constructed object or artefact, or perhaps a piece of engineering or an archaeological 'remnant' or 'ruin', etc). In the latter case, in establishing the existence of a *corporeal* sign (e.g., a physical trace, object, artefact) we may also have made use of *electromagnetic* signals which have brought the information about such an object or trace to us (such as, for example, telescopic imaging of a spacecraft in a distant part of the solar system), but, as should be clear, it is *not* the electromagnetic signal *itself* which is the sign—it is merely the carrier of information about the *actual* physical sign.

Another aspect of the sign is whether its *character* is deliberate and intentional, or simply unintentional and incidental. The radio and optical SETI searches have been looking for both—whether a deliberate signal, beamed intentionally at Earth, or simply incidental 'leakage' of, so to speak, 'domestic' radio or optical transmissions of that technological civilisation, much as our own television broadcasts are an unintentional side-effect of our civilisation and could become incidental signs to any nearby civilisations who might be conducting SETI searches themselves. Hence, let us choose two main values for the *character* of the sign: *intentional*; and *incidental*.

### 3.2 The Contact itself

As may be clear, one of the more important parameters characterising contact would be its *proximity* to Earth, especially as regards the potential response of humanity *to* such contact. This can be considered for our purposes here as having six useful practical values: terrestrial (i.e., taking place on Earth); peri-terrestrial (i.e., taking place 'near' the Earth); in the rest of the Solar System as a whole; in the solar neighbourhood (i.e., 'nearby' stars); in the wider Milky Way Galaxy; and, lastly, extragalactic (i.e., outside the Milky Way Galaxy). One can easily see how this value could in fact condition the human societal response to a very great extent. A terrestrial discovery of a clearly extra-terrestrial intelligently-constructed artefact would likely generate a much bigger societal response than, say, the discovery of an ancient fossilised simple bacterium in a rock on Mars, although—perhaps because it *was* a terrestrial (i.e., highly proximal) discovery—the famous case of the Antarctic Martian-origin meteorite ALH84001 [50] demonstrates the intense underlying public interest in this topic, which even prompted the then US President Bill Clinton to make an official announcement [54].

There is also merit in at times explicitly considering a somewhat subtler distinction having to do with the form or *class* of contact, namely whether it is 'direct' or 'indirect'. The 'directness' or 'indirectness' of any actual contact may also, like proximity, produce somewhat varying levels of potential response. One can imagine cases where 'contact' with an entity is quite indirect, such as inferring biological life on nearby exoplanets via interpretation of the chemical composition of the atmosphere through spectroscopy undertaken on the light coming from the planet. In such an instance, we are obviously not in contact 'directly' with the entity. An example of 'direct' contact could be the actual finding of extant bacterial life on Mars or the moons of Jupiter or Saturn, or, indeed, the receipt of a clearly-deliberate and carefully-directed "Hello Earth!"-type radio signal from a nearby star system. One might choose to characterise finding actual physical, albeit merely fossilised, evidence of an entity as either direct or indirect, depending on one's view of what constitutes 'directness'. The subtleness of this parameter allows for some useful nuance in characterising the type of contact, and the likely response which that contact might engender.

### *3.3 The ‘Entity’*

Another significant parameter is the *complexity* of the ‘entity’ with whom/which contact occurs. Anticipating the following discussion, and paying mind to the earlier comments regarding the need to extend the initial assumptions of SETI regarding biological beings to include consideration of possible *post*-biological intelligence, we do not want to implicitly assume or presume upon the physical form or nature of the ‘entity’ with this parameter. Rather, we would like to reserve the possibility of making contact with ‘entities’ that may or may not be biological in nature, and may or may not be intelligent, so we need a way to characterise their complexity that is independent of any assumed physical form. Accordingly, utilising commonly used terms from the SETI literature, the *complexity* parameter of the ‘entity’ will be assumed to have three main values: *simple*; *complex*; and *intelligent*. These are sufficiently general that they need not assume a biological basis, and are also deliberately left somewhat indistinct as to the exact dividing lines between them, as we are attempting to encapsulate, using only a few main values, what is actually a broad continuum, so some conceptual overlap is both useful and helpful here. Thus, in the case that these are indeed biological entities, then the two-parameter combination *simple-biological* could characterise single-celled microbial life, while *complex-biological* might characterise large multi-cellular organisms such as reptiles, and *intelligent-biological* would suitably characterise human beings.

Given the foregoing, another important parameter therefore is the form or *nature* of the ‘entity’ itself. As suggested, there is some careful consideration that needs to be given to this. One value would of course be ‘biological’, as we are an instance of this ourselves (although precisely defining biological life is not so obviously straightforward as one might hope or imagine [29]). Another would be ‘technological’ (i.e., a machine), while yet another relevant plausible value based on our current state of knowledge would be a ‘hybrid’ of technology and biology (however the concept of ‘biology’ is conceived). This is usually denoted in both the futurist and SETI literatures as a ‘cyborg’ [e.g., 28, 47], although time-scale arguments suggest that this could well be only a relatively short-lived ‘transitional’ form between biology and a fully post-biological form of entity [e.g., 25]. Hence, another way to express the value ‘technological’ might also be as ‘post-biological’, given earlier comments along these lines, although we may decide in some analyses to reserve this latter value as somewhat separate from the value ‘technological’ as one can imagine sentient machines that are a form of post-biological super-intelligence which nonetheless makes use of ordinary non- or quasi-intelligent complex machines for the purposes of exploration (much as we do now with our space probes). Therefore, for our initial analysis here, the values of the *nature* of the *entity* are chosen to be: *biological*; *hybrid/cyborg*; and *technological/post-biological*, with the option of unbundling the last one when it seems appropriate.

Finally, since energy is the most essential fundamental currency underpinning the functioning of the universe, we should also consider the entity’s *Kardashev Type*—the magnitude of energy utilised by the entity as part of its form of social organisation (if indeed it possesses one). Kardashev’s classification was based on the amount of energy which is used or can be controlled by technological civilisations [48], and, in brief, the Kardashev schema defines civilisations through their energy use as follows [e.g., 70]:

**Type I: planetary.** A Type I civilisation is one which is able to make use of use all of the available energy of its home planet, estimated to be on the order of $10^{16}$ watts or 10 PW. This would include harnessing, for example, tidal, thermal, atmospheric, nuclear, fossil, internal and other planetary sources of energy.

**Type II: stellar.** A Type II civilisation is one which has managed to harness all of the energy output of its home star, something like $10^{26}$ W or 100 YW. This might include collecting all of the radiant energy of the star, and/or perhaps even harnessing the energy contained in its gravitational field. Such a civilisation might even be detectable over intra-galactic distances (c.f, e.g., Dyson [31]).

**Type III: galactic.** A Type III civilisation is one which has managed to harness the energy of an entire galaxy, something like $10^{36}$ W, although this figure is somewhat variable because galaxies can vary considerably in size. A civilisation capable of using energy at this scale could probably make itself visible, if it chose to, over inter-galactic distances, including perhaps throughout a sizeable volume of the observable universe.

The original schema has also been refined by many others over the ensuing decades and extended to include a decimal interpolation between values, the value 'zero' and even 'negative' values, as well as values beyond 'III' [9, 15, 70]. In this view, Earth is not yet considered as being at Type I and so is generally referred to as a Type '0' civilisation (indeed, Type ≈0.72 by some recent estimates [e.g., 15]). We can, following [9], use the schema to characterise the energy use of any type of entity, absent any assumption of there being a 'civilisation' or even intelligence involved, which usage might even utilise negative values for, say, colonies of bacteria. For simplicity of the analysis, let us simply use '0' to denote anything up to and including Type 0, and accordingly use the types: 0, I, II, and III as the principal parameter values, delaying any further extension until possible subsequent exploration with a more-expanded or different parameter set than is used here.

### *3.4 Other possible additions to the parameter space*

The three broad parameter categories used above could of course each have several more parameters added to them, a few of which we shall note in passing as possible extensions for future work, but not pursue any further here.

It would be possible, as foreshadowed above, to extend the 'sign' category of parameters to expand upon the current simple two-part distinction between what amounts to an immaterial 'signal' carried by quantum particle-waves (here assumed to be electromagnetic) and some kind of corporeal physical manifestation, which latter could be expanded into a distinct sub-space explicitly distinguishing between the entity itself, a fossil, an artefact, architecture, engineering, an archaeological trace or remnant etc. For the 'signal', the choice of frequencies to use (radio, optical, X-ray, gamma-ray, etc), and even its quantum-particle composition (expanding the current case of electromagnetic 'photons' to include gravitons, neutrinos, and indeed anything else we might discover…) could also be individually considered in detail. In this case, we would change the overall parametric designator from 'electromagnetic' to something suitably general, such as *quantum-particulate*, for example, for which 'electromagnetic' would then simply be one of the possible parameter values (recall the 'contingency thinking' discussion from Section 2). Here, for our purposes, it suffices to treat these parameters at the higher level of generality we have used in order to demonstrate the utility of the approach without drowning in too much (albeit potentially-fascinating) detail. The advantage of this sort of technique is that, once specific parametric configurations have been identified, it would be possible to 'zoom in' to unpack the finer details of possible, as it were, parametric 'sub-spaces' that might lie further 'within' a particular configuration or 'general' parametric value. Bulk manipulations of the totality of the morphological space are probably best carried out at a higher level of granularity, and the finer level reserved for particularly interesting configurations found in the more granular possibility space. (This type of more detailed analysis of multiple and nested sub-spaces would benefit greatly from suitably-sophisticated computer software; see the comments in Section 9.)

Another possible parameter for the 'entity' category might be '*motivation*'—such as: *indifferent, benign, helpful, belligerent, malevolent* [e.g., 43, 45]. We could also consider the possible relationship the entity might have to us: *safe, hazardous*. This latter parameter could represent both the 'hazard' represented by the physiology of a simple microbial organism as well as perhaps by, as it were, the 'psychology' (if such a term is even fitting when applied to post-biological extra-terrestrials) of an intelligent entity. A simple microbial organism would likely have 'indifferent' (i.e., be absent any conscious) motivation, but might be hazardous from a physiological perspective. This consideration was of course at the basis for the biological isolation protocols of the US Apollo space program wherein the Apollo XI astronauts were placed into temporary isolation upon their return to Earth. In popular culture, it was also the basic plot element of the 1969 novel and 1971 film *The Andromeda Strain*, and, as popular culture more than abounds with examples of hazards posed to Earth due to extra-terrestrial 'psychology', we need not dwell any longer in noting this point.

Another category entirely of parameter in the possibility space—which has quite strong ties to contemporary SETI discussions related to the 'entity-motivation' parameter—could be our own

'stance' with respect to SETL or SETI: *passive*; or *active*; which yields both conventional and 'Dysonian' SETI for the first value, as well as the more 'active' variant known as 'active SETI' or 'METI' (*Messaging* to Extra-Terrestrial Intelligence) for the second. This more active 'stance' could conceivably influence the timing of any potential contact to some degree, although proposals to do so are considered quite controversial, having strongly-argued positions both pro and con [see, e.g., 44], owing to the risky possibility that an *intelligent, proximal* (e.g., solar neighbourhood) entity might also be *belligerent/malevolent-hazardous*. The suggestion for SETI to remain in the passive stance to avoid this possibly-existential risk is sometimes referred to in the SETI literature as the prudence of 'refraining from "shouting out into the jungle" before we know what is out there' [e.g., 5, p. 163]. Of course, some of our past attempts and future proposals to search for life in the Solar System by launching specialised probes represent an 'active' form of SETL, but these are usually considered somewhat less risky than METI.

These, and many other possibilities [cf., e.g., 89], are all very interesting parameters to consider in any future more comprehensive work, but for our present purposes—to map out a *preliminary* working schema for examining the possibility space for a discovery during SETL or SETI—they are less essential. Later and subsequent analyses could focus on more specific possibilities and extensions, a few of which we have noted above. But for now let us turn to use this schema to serve as the basis for beginning to think systematically about the scenario space for contact.

## 4. The Sector/Factor Array for Contact – 'LSEARCH'

We are now ready to construct the Sector/Factor array for this particular choice of parameters as the basis for our preliminary morphological analysis of the scenario space of contact.

The seven parameters/Sectors chosen as a basis for analysis in the preceding section can be summarised as in Figure 1, where (per the FAR approach to morphological analysis) a particular letter from that parameter name has been highlighted for use as the Sector designator. This shows graphically the scope of the parametric space being considered here.

[Take in Figure 1 about here]

**Figure 1**: Initial characterisation of a preliminary parametric space for contact

From noting that Sectors S, A, and C each have 2 Factors, Sectors L and E each have 3, Sector H has 4, and Sector R has 6 Factors, we can quickly calculate that this Sector/Factor array represents $2\times2\times6\times2\times3\times3\times4 = 1728$ ($=12^3$) possible contact scenarios (absent any potential reduction due to formal cross-consistency assessments, non-coherence filtering, or anomaly elimination).

Let us now, however, construct a more evocative mnemonic 'word' that will simultaneously remind us of the full set of Sectors we are using, as well as allow us to discuss specific Sectors and their Factors unambiguously. Obviously, the more memorable the mnemonic, the better. Hence, with a bit of rearranging of the Sector names, and recalling that the actual order of letters in the FAR approach is immaterial except for the sake of convenience of pronunciation, we might decide to choose the rather appropriate term 'LSEARCH', pronounced 'ell-search', as our 'word'. Figure 2 shows both the Sector names in their new order, as well as the Factors and their letter-numeral designators; thus, the second Factor of the comp**L**exity Sector—a 'complex' entity—may be written equivalently as "L2", "L2", or "$L_2$", and we will make use of these comparable denotations in what follows.

[Take in Figure 2 about here]

**Figure 2**: The LSEARCH Sector/Factor array; a minimalist morphology for contact scenarios

As we prepare to move into the detailed analysis, it is appropriate to introduce a form of notation which will prove convenient in the following (this notational form was earlier described briefly in [93]). As we will be concerned with examining sub-spaces of the morphological configuration space in addition to denoting specific parameter/Sector values/Factors, such as just shown for "$L_2$", we may also find it convenient to simply denote a Sector in general with no particular values/Factors specified. There are two ways we do this. In the first, when the specific Factor values of a Sector are 'unspecified' and not of direct relevance to the discussion, we will simply use the Sector letter by itself without any index marker; so Sector L in general would be denoted simply "L". In the second, when the Factor values of a Sector *are* being explicitly considered as part of the discussion but are also left unspecified while being left 'free' to take on any value, they are marked as such with a *text* index rather than a *numerical* one; this would be rendered as "$L_j$". In this latter case we will also use different index letters on the different Sectors in a single denotation, as a way to further emphasise notationally the independent nature of the different Factor values in the different Sectors which are being explicitly considered but which are otherwise being left 'free'. Thus, for the sake of illustration, the explicit consideration of both the complexity and proximity parameters, and no others, without specifying any particular value of either parameter, would be denoted "$L_iSEAR_jCH$" (or sometimes perhaps the even simpler form $L_iR_j$ where this is more convenient). In this notation, the specific *index* letters have no significance at all (as opposed to the Sector letters which of course do), and are simply placeholder indicators, in the same way that free index letters in tensor calculus are similarly used merely as placeholders where necessary [e.g., 11].

With these preliminaries, let us now use this schema as the basis for our further explorations.

## 5. Identifying approaches to SETL and SETI in LSEARCH

A notable proposal to classify contact scenarios—and which was one of the inspirations a decade ago for the initial discussion in [92] and the follow-up exploration being reported here—was the one due to (the late) Albert Harrison [42, 45], based on a simplified form of the complexity-proximity subspace $L_iSEAR_jCH$ (or $L_iR_j$). As noted in [2, 93], when the number of parameters $n = 2$, and each parameter has only two values, $k_1 = k_2 = 2$, one then recovers the more well-known 2×2 matrix form of scenarios used extensively by some futurists [e.g., 87]. In the scheme Harrison used, L had the two values *simple* and *complex* (combining the main foci of interest for SETL and SETI, respectively, where *complex* in Harrison's usage is essentially equivalent to what we have here termed *intelligent*), and R had the two rather general values *proximal* and *distal.* Thus, the four contact scenarios which emerged were: *Microbes on Mars* (simple-proximal); *Distant Dust* (simple-distal); *ET Calling* (complex-distal); and *Space Visitors* (complex-proximal), which also shows the use of the sort of evocative scenario names that are so beloved by futurists. These were and still remain very useful first-approximation archetypes for thinking about the possibilities for contact when the parameter space is confined to a smaller 2×2 subspace of $L_iR_j$, although it would be possible to immediately extend this classification very easily to examine the (3×6=) 18 scenarios entailed by the extra values for L and R used here.

One of the things a morphological approach attempts to achieve is exhaustiveness, so we should be able to recover our own case from the LSEARCH array, as a first-run 'reality check' on the completeness of the morphological space. That is to say, if we ourselves have been 'discovered' by a nearby extra-terrestrial technological civilisation previously unaware of us, how would the scenario of *their* detection of *us* (which we might call "*Discovery of Earth*") be characterised in this schema? Probably like this: $L_3S_1E_1A_2R_4C_2H_{0/1}$—namely, a 'nearby' intelligent, biological entity, at a pre-planetary stage of civilisation (although perhaps heading towards Type I), detected through incidental 'cultural leakage' electromagnetic signals, $S_1A_2$ (as we do not—yet—undertake an 'active SETI' program, $S_1A_1$), and where no 'direct' contact has been made with the entity (i.e., us), hence $C_2$.

Our own case having been successfully recovered, let us now try to characterise some of the existing and historical approaches to SETL and SETI using the LSEARCH schema.

Early conventional SETI (i.e., mid-20th Century CE, such as Frank Drake's famous pioneering Project Ozma which looked at two nearby Sun-like stars [28]) could be characterised by the denotation: $L_3S_1E_1A_iR_4C_jH_{0/1}$—which is to say, very similar to the configuration of the *Discovery of Earth* scenario just considered, except extended to include *both* index values of Sectors A and C. This denotation represents a search for *either* deliberate *or* incidental electromagnetic signals from nearby civilisations that may be comparable to us (i.e., intelligent, biological, Type 0 or perhaps I, and with the built-in assumption of being situated around a 'suitable' habitable Sun-like star), and allowing for the possibility of both 'indirect' and 'direct' contact with them. This type of search continues to this day, albeit with expansions of the search space by extending the electromagnetic frequencies used to include optical wavelengths, and expanding upon the spectral class of candidate stars [35, 74, 85, 86].

As mentioned earlier, the initial SETI search strategy was prompted in part by the 1959 Cocconi and Morrison paper [26], which noted that it would be possible to detect civilisations comparable to us using our then-existing radio telescope technology [16], and we can clearly see this in the close similarity of *Discovery of Earth* and Project Ozma. In short order, as was also noted, Dyson stepped in with his suggestion to instead look for signs of alien *technology* [31, 32], which proposal gave rise to the famous concept of the Dyson Sphere/Shell [4]. His suggestion can be characterised by the denotation: $L_3S_2EA_2R_{4/5}CH_2$ (here ignoring the specific nature of the entity and the class of contact). The contrast between conventional 'orthodox' SETI and 'Dysonian' SETI may thereby be characterised most simply by the distinction between $S_1$ and $S_2$, respectively, as well as by the general underlying assumption that any observable technology or engineering activity was not necessarily created or undertaken to purposefully attract attention—the Dysonian approach thus focussing principally on $A_2$.[1]

In contrast to SETI, where the focus is on $L_3$, astrobiology and SETL are more usually concerned with $L_1$, and possibly $L_2$. Proposed missions to search for evidence of life on Jupiter's moon Europa and/or Saturn's moons Enceladus and Titan would have designation $L_{1/2}S_2E_1A_2R_3C_jH_0$ (i.e., either $C_1$ direct currently-extant 'living' or $C_2$ indirect ancient 'fossil' evidence of simple microbial $L_1$ or perhaps even more complex $L_2$ life). We might decide to give Mars an $R_2$ designation ('near' to Earth in the Solar System), in contrast to $R_3$ for the more-distant gas giants and their satellites, so several NASA missions to Mars (e.g., the Viking landers of the 1970s, and the more recent rover missions) might be characterised by the very similar designation $L_{1/2}S_2E_1A_2R_2C_jH_0$. In this view, the discovery of any still-extant bacterial life on Mars would have the denotation $L_1S_2E_1A_2R_2C_1H_0$, while the discovery of a rock containing a fossilised specimen of an ancient bacterium would also fall under this denotation with the contact class C instead regarded as $C_2$ (i.e., 'indirect'), with both cases being versions of Harrison's *Microbes on Mars* scenario (and with the former more likely to engender a somewhat bigger human societal response than the latter). The famous controversial case of the Antarctic meteorite ALH84001 mentioned previously—had it turned out to have indeed been such evidence—would thus have had an LSEARCH denotation of $L_1S_2E_1A_2R_1C_2H_0$, thereby differing only in the index value of R.

The Kepler mission launched late last decade to search for Earth-like exoplanets in the Milky Way Galaxy can also be characterised in this schema—it is a way to expand the search space for possible habitats for extra-terrestrial life from $R_{2/3}$ to $R_{4/5}$. One scenario for contact emanating from thinking about the Kepler mission findings would, as Shostak suggests [76], be detecting chemical signatures for biological activity in the atmosphere of an exoplanet in the spectrum of the light captured from imaging that planet. This circumstance would have an LSEARCH denotation of $LS_2E_1A_2R_{4/5}C_2H$, where the actual complexity L and Kardashev Type H of the entity have not (yet) been explicitly classified, and where we have chosen $S_2$ rather than $S_1$ because we would have detected physical/chemical evidence of biological activity (i.e., even though this information has been carried to us by way of electromagnetic waves, these waves are not themselves the actual 'signature' of life, such as an obviously-intelligent electromagnetic radio signal would be). If the atmospheric gases so detected are consistent with what we would consider to be plausible naturally-occurring basic biological processes, we might tentatively propose the further designations $L_{1/2}$ and $H_0$. However, if the gases inferred from such a spectral analysis instead appear to be consistent with what we might consider products of non-natural quasi-industrial processes, we might instead assign the designations $L_3$ and $H_{0/1}$ and immediately bring our radio telescopes to bear in order to look for any possible $S_1A_j$ (deliberate or incidental electromagnetic) signals.

[1] Interestingly—and this arises from the morphological perspective taken—it appears that there may be a significant correlation between proximity R and Kardashev Type H in this schema; that is, as the proximity parameter increases to greater distances, so too would the Kardashev Type of the entity need to increase, if a cross-consistency criterion of 'plausible detectability using our current level of technology' is now applied. This is an example of where not applying cross-consistency assessments *a priori* may be able to generate further insights from considering such parametric 'clashes' as they might arise *post hoc* during more detailed exploration and analysis.

Let us now examine a view noted earlier due to Davies and Wagner [23], who suggested looking on the Earth's Moon for signs of alien artefacts. This would have the LSEARCH contact designation of $L_3S_2EA_jR_2C_2H$, while related suggestions to look in the asteroid belt, Kuiper belt or even Oort Cloud for signs of activity [e.g., 33, 56] could be correspondingly rendered as $L_3S_2EA_jR_3C_2H$, differing only in the index value of R. We have chosen A with both index values since any artefacts or evidence of activity could have been left behind either incidentally (e.g., as junk, or as the residue of mining or resource-extraction activities) or intentionally (e.g., as a deliberate sign for anyone who might come along later).

A very interesting suggestion that also bears mentioning—due to its initial apparent simplicity actually hiding a very fruitful line of possible future exploration—is the one discussed by the late Allen Tough, who noted the possibility of our encountering a smart interstellar probe launched by another intelligent species long ago and which has now found its way into our solar system [83]. (This scenario is clearly inspired by our own recently acquired ability to launch a probe beyond the boundaries of our own solar system, extrapolating from our present technological base into the distant future, and imagining if other intelligent technological species may have also done so before us.) There are, as it turns out, many, many possible LSEARCH designations that could emerge from expanding upon the thinking around this situation, and teasing out the subtle nuances makes for some quite agreeable inquiry. For example, what is the 'entity' here—the actual proximal machine, or some putative more-distant creator-entity? This will clearly affect the values for L and E in any characterisation. Was the contact the result of an $A_1$ intentional directed mission or simply an $A_2$ incidental outcome of more general undirected exploration? There are also many possible characterisations that could arise depending on how much we are willing to infer or speculate on the L and E (or even H) values of any separately-analysed supposed creator-entity, as opposed to the L and E values of the physical artefact/object/machine before us (e.g., this could involve the use of a kind of extended two-part LSEARCH characterisation of contact, one for the physical machine/artefact and one for the assumed creator-entity). Or, for that matter, what if it turns out that the 'machine' *is itself* an instance of post-biological intelligence that is unbound to planetary or even circumstellar existence and has chosen to travel from star system to star system in order to explore and 'see the Galaxy'? This would also alter the character of the contact scenario quite markedly (and the likely human response, too). Given all this, we might decide to denote this scenario sub-space with the rather 'sparse' denotation $LS_2EAR_{1\text{-}3}CH$, leaving unspecified all index values except where we found the corporeal object, whatever its ultimate nature turns out to be. This noteworthy example of a futurist thinking about SETI—cf. Frank Drake's comments in the Introduction—shows at least one possible line of future refinement of this schema, and many possible directions for future investigation. Indeed, a detailed exploration of the possibilities lying within the $L_3E_3$ (intelligent post-biological) sub-space would be a fascinating extension to this analysis (one brief example of which can be found in [96]).

Finally, another particularly interesting related proposition that also warrants mentioning here is the recent "starivore/stellivore" (i.e., star-eating) hypothesis of Vidal [88, 90], which posits—and argues extensively on the basis of thermodynamic, computing and several other criteria—that suitably-advanced post-biological Type II civilisations would end up not merely passively utilising the energy of, but rather *actively feeding upon*, their star, and would have the same general appearance to us as what are known in astronomy as 'accreting binary systems'. He proposes a number of tests to distinguish between what are naturally-occurring and what might be candidate putatively-artificial binary star systems. In terms of the LSEARCH schema, we might characterise this intriguing proposal as $L_3S_2E_3A_2R_5C_2H_2$, and it will be most interesting to see how any further more direct investigation of this fascinating hypothesis pans out…

The reader can now see (I hope) how this schema can be employed to usefully classify approaches to SETL and SETI and therefore to classify potentially successful detection scenario(s) that may arise from those activities. It would be possible to go into even more examples—and those who are steeped more completely in the SETL and SETI literature or enterprise might wish to do so for the sake of further exploration—but this sampling should be sufficient to establish the basic method and demonstrate the thinking used to employ it.

We now briefly consider some of the many fictional scenarios for contact as they have been depicted in popular culture.

## 6. Identifying fictional contact scenarios in LSEARCH

Popular culture is full of (so far!) fictional scenarios that depict the circumstance of our encountering evidence of life off the Earth, and we can only hint at the totality of the large number of such depictions. Hence, we will here only focus on a very few of the more seriously plausible of these.

The archetypal plausible serious contact scenario ever since the advent of SETI has been the (*ET Calling* [42, 45]) 'signal of intelligent origin from space' scenario depicted in the 1997 film *Contact*, based on the best-selling novel of the same name by Carl Sagan [71], one of the pioneers of SETI (and to whom the film is dedicated). This scenario is clearly $L_3S_1EA_1R_4C_1H$. In the film, news of contact is widespread and the resultant human societal response ranges from joy and rapture to terror and dread. It represents one of the few rare instances of an intelligent quasi-realistic Hollywood depiction of contact, as opposed to the more usual block-buster 'space opera' or 'space war' tropes—although the one well-documented instance of a signal that genuinely looked for a while like it might *actually* be 'it' [75], did not engender either the speed or magnitude of societal response depicted in the film.

Another earlier and similarly intelligent depiction is the one from the 1968 film *2001: A Space Odyssey*, co-written by noted SF author Arthur C. Clarke and the film's director Stanley Kubrick. In this scenario, exploration of the Moon uncovers an ancient artefact of clearly-artificial origin which subsequently sends a radio beam in the direction of Jupiter, prompting a mission to investigate further, where a similar but much larger artefact is located. This scenario is initially $L_3S_2EA_1R_2C_2H$ (cf. [23]), as the artefact is dug up on the Moon by American scientists, but becomes $L_3S_2EA_1R_{3\text{-}5}C_{1/2}H_{1/2}$ as it becomes clear that the system transporting the astronaut at the film's end takes him out of the solar system entirely to another planet and must represent an, at the very least, Type I and probably Type II level of technology. In this film, in contrast to *Contact*, the initial discovery is kept an official *absolute secret*, including from other governments, with the desire to do so leading to unintended consequences for the subsequent mission to Jupiter.

Other more recent popular films have mostly had a tendency to focus on $R_1$ proximity, the action generally tending to involve (usually) biological aliens with a *malevolent* motivation to attack, capture or destroy the Earth. This sub-genre has a long history, dating back in literature at least to H.G. Wells' 1898 novel *The War of the Worlds* (where the entities were biological Martians $L_3E_1R_2H_0$), and the subsequent 1938 adaptation as a radio drama by Orson Welles, as well as its adaptation to at least two major films (1953, 2005). These days, it seems every year a new blockbuster or two comes out with some new $R_1$ existential threat to humanity from malevolent extra-terrestrials, so the reader will likely be well-acquainted with these. Notable exceptions to this plotline have been *The Day the Earth Stood Still* (1951), *Close Encounters of the Third Kind* (1977) and, more recently, *Arrival* (2016), with each depicting the arrival on Earth of biological extra-terrestrials ($L_3S_2E_1A_1R_1C_1H$) with a generally benign or possibly even benevolent motivation, although—as in *Contact*—the human societal response is quite varied.[2]

Cinema is constrained by the need to portray narrative visually. Literature, on the other hand, allows much more free reign for the imagination, so this genre is a much richer source of potential contact scenarios. As a result, one can find all manner of the LSEARCH parameter values represented in the science-fiction literature, and the reader is invited to explore and classify his or her own favourite examples using this schema.

It is also possible to randomly vary parameter values at will in order to, as it were, 'see what scenarios show up' by reverse-engineering what the randomly-generated parametric designation might look like or map to in the reality it is intended to describe, a technique we will briefly discuss in the next section.

## 7. Random generation of configurations for further exploration

Owing to the agglutinative nature of the process of constructing the morphological configuration space from a diverse set of largely 'orthogonal' multivariate parameters, one of the things one can do 'for fun', especially in a large morphological space, is to, as it were, 'spin the Sectors' like the slots on a poker machine to see what multi-Factor 'hand' is so dealt, and investigate the resulting configuration(s) further. In this approach one allows for the Sectors to take unspecified or 'free' values, which we can variously denote with just the Sector letter itself; an index letter rather than a numeral; or even with multiple values not including all of the possible values, by using multiple index numerals. This provides a mechanism to generate a random configuration for further exploration, quite apart from the more systematic and exhaustive exploration which a morphological approach affords, and it is especially interesting to do this when entering into the new unknown territory of a recently-constructed morphological space. This 'slot machine' technique was employed to generate scenarios for use in a recent research project examining potential R&D priorities for the future of the Australian building and construction industry [3, 72, 98].

In late 2009, around the time [93] was published, the author revisited the discussion in [92] concerning the parameter space of contact and—on daily long train commutes—began experimenting with various random configurations of the LSEARCH parameter space 'just for fun' to 'see what possible contact scenarios showed up'. One of the configurations that arose in this way is as follows: $L_3S_2E_iA_jR_6C_2H_{2/3}$ (i.e., with unspecified 'free' values of E and A). On closer examination, this configuration sub-space in fact represents a very intriguing set of possibilities—what we might characterise as *galaxy-scale macro-engineering* taking place outside the Milky Way Galaxy, undertaken by intelligent entities which have reached at least Kardashev Type II and may be heading towards or have possibly reached Type III. This configuration was so intriguing that it has persisted in mind continuously over the intervening years. Ultimately, it became part of formal lectures to post-graduate students later in 2010 (on macro-perspectives beyond Earth, based on [92]), the subject of several conference presentations in subsequent years [94, 95, 97], and was eventually written up more fully for the selected-papers volume of a 2012 conference on Big History [96]. It is so intriguing precisely because it appears, at least *prima facie*, that there *might* exist what could possibly amount to a real example of the situation represented—namely, the beautiful ring galaxy known as 'Hoag's Object' (see Figure 3).

[Take in Figure 3 about here]

**Figure 3:** Hoag's Object (PGC54559).[3]
Image credit: NASA & The Hubble Heritage Team (STScI/AURA). Acknowledgement: Ray A. Lucas (STScI/AURA).

---

[2] By contrast, in the 2008 remake of *The Day the Earth Stood Still*, the initial motivation towards humanity was rather less benign and rather more *indifferent.*

[3] Multiple formats and sizes are available from: http://hubblesite.org/image/1241/news_release/2002-21.

## 8. In brief: the intriguing case of $L_3S_2E_iA_jR_6C_2H_{2/3}$ – Hoag's Object?

Space does not permit a full discussion of this enchanting possibility here, although a more complete exploration and plausibility argument can be found in [96], as can some proposed empirical observations that could be made to test the hypothesis. Suffice it to say, that the remarkable structure of Hoag's Object (which has the formal designation of PGC54559 in the Principal Galaxies Catalogue), has attracted some interested attention since its discovery in 1950 by Arthur Hoag [46]. A number of papers exploring the possible nature, structure and evolution of Hoag's Object have since been written [7, 8, 36, 40, 55, 58, 73], with the explanations for its unusual morphology having principally focussed on attempting to model naturally-occurring dynamics as the underlying mechanism. Most of these proposals remain entirely viable hypotheses, of course, and there is no intention here to either criticise or cast any doubt upon their potential viability. Rather, primed with the LSEARCH designation $L_3S_2E_iA_jR_6C_2H_{2/3}$ in mind, and the discussion and thinking that have led up to it, the reader is invited to consider the image shown in Figure 3 in this new light. The striking 'core-gap-ring' morphology seems to be very suggestive of at least the possibility that, just perhaps, some other-than-entirely-natural processes might have been at work… At the very least, it would seem to be such a captivating possibility that taking a careful closer look might be warranted, 'just in case'…

Three specific empirical observations that could be undertaken to do this are [96]: (i) examination of the radiation coming from behind the galaxy through the 'gap' between the core and ring for any anomalies (e.g., diffraction, scattering, polarisation, etc) compared to analogous background radiation from immediately adjacent to it; (ii) spectroscopic examination of the composition of the ring for any anomalies in the extent or character of star-forming regions or in the chemical composition (e.g., metallicity profile) compared to what would be expected from the 'usual' processes of stellar evolution going on in analogous spiral galaxies of similar diameter and age; and (iii) examination of the peri-core region for any possible time-keeping or other beacons that may be being used to coordinate in time any putative galaxy-scale engineering activities that might require synchronisation. These types of observations are intended to provide specific concrete empirical tests of the hypothesis contained in the (fairly beguiling!) research question: *Is Hoag's Object a Dysonian artefact?* If the answer turns out to be in the affirmative, then this would be a contact scenario which, due to its extremely distal nature (with $R_6$ ~600 million light-years), might be able to provide a confirmation of $L_3$ without necessarily engendering some of the more fearful human societal responses that other more proximal contact scenarios with the same entity-complexity value might elicit.

## 9. Concluding remarks: Toward scenarios of the human societal response to 'contact'

Of course, *actual* contact would be an enormously significant event in the ~250,000-year history of the human species [12, 78]—with potentially huge implications for our conceptions of ourselves and our place in the Universe—and could be a trigger for what might be a quite profound 'challenge' [51] to our present-day civilisation. As Toynbee noted in his model of macrohistory [37], every historical civilisational challenge has prompted the need for some form of response, and it is the character and quality of that response that has been of profound importance for the future of that civilisation.

Here we have focussed on the 'contact'/'challenge' side of the event. A separate morphological analysis covering a 'response' that is appropriately 'matched' to the 'challenge' is something for future work. The approaches of other researchers of 'matching' one morphological space to another was discussed in some detail in [93], and it would be very valuable to draw upon these various techniques in extending the exploration undertaken here. Clearly, and as has been intimated here in several places, a potentially large number and variety of parameters could be involved in characterising our possible societal or civilisational 'response' to the 'challenge' of contact itself. The recently-delineated 26-parameter impact model of Vidal [89] would be an excellent starting point for further elaboration of this parameter space, as would the other earlier work which has for decades been considering this question [e.g., 1, 24, 42, 45, 79, 83]. It would also seem that in order to allow for easier exploration of such greatly-expanded parametric representations, the use of appropriately-sophisticated software allowing for many more parameters and for parametric sub-spaces to be hidden or expanded as needed would also be required [e.g., 65, 66], if any serious attempt at comprehensiveness is to be made.

It is fitting to end this preliminary exploration of one possible morphology for contact scenario space here by re-iterating a comment by Albert Harrison and Steven Dick made while surveying the four scenarios arising from their use of the 2×2 subspace of $L_iR_j$ discussed in Section 5:

> This framework was designed as a useful heuristic for organizing various hypotheses about long-term and short-term reactions to [extra-terrestrial][4] life. It should be useful also for developing scenario-contingent strategies for managing contact and its aftermath [45, p.14].

The same underlying sentiment applies to the more expanded LSEARCH schema presented and described here. Our restricted choice of just seven principal parameters was made in order to cover the scenario space as broadly as possible at a high level of granularity while keeping the analysis tractable in this initial exploration of this potentially quite vast area of inquiry. Although a particular specific choice of parameters was made here, for the sake of definiteness of presentation, the approach taken can clearly be generalised to include many other and many more parameters, if desired. It would be very welcome indeed, if the approach sketched out in this paper were modified and extended over time by other interested researchers as our collective thinking progresses, to expand upon the current analysis using the systematic contingency thinking that the morphological method helps to both generate and support.

In fact, it is the author's fond hope that precisely such further extension, elaboration and exploration occurs, and that the method is taken up and adapted by other researchers who are also interested in seeking to undertake a more extensive and exhaustive systematic exploration—both of the possibility space of potential contact scenarios, as well as of the range of possible human societal responses prompted by what would be, frankly, a truly momentous and unprecedented occurrence in human history. And, to that end, it is also fondly to be hoped that an empirical observing program of Hoag's Object be undertaken to search for any subtle signs of possibly-artificial activities, given the astonishing pay-off that such a modest investment might just yield…

□

[4] In the original, the word 'human' is used, although from the context this is clearly a printing error since, when compared with the earlier workshop paper abstract of Harrison [42], the word 'extraterrestrial' is found in the same position.

| Sign | | Contact | | Entity | | |
|---|---|---|---|---|---|---|
| compo**S**ition | ch**A**racter | p**R**oximity | **C**lass | comp**L**exity | natur**E** | kardas**H**ev type |
| electromagnetic | intentional / deliberate | terrestrial | direct | simple | biological | 0 |
| corporeal / physical | unintentional / incidental | near-Earth / peri-terrestrial | indirect | complex | machine / biology hybrid (cyborg) | I |
| | | solar system | | intelligent | technological / post-biological | II |
| | | solar neighbourhood | | | | III |
| | | wider Milky Way Galaxy | | | | |
| | | extra-galactic | | | | |

**Figure 1**: Initial characterisation of a preliminary parametric space for contact

| L<br>entity compLexity | S<br>Sign compoSition | E<br>Entity naturE | A<br>sign chAracter | R<br>contact pRoximity | C<br>Contact Class | H<br>entity KardasHev type |
|---|---|---|---|---|---|---|
| **L1**: simple | **S1**: electromagnetic | **E1**: biological | **A1**: intentional | **R1**: terrestrial | **C1**: direct | **H0**: Type O |
| **L2**: complex | **S2**: corporeal | **E2**: hybrid / cyborg | **A2**: incidental | **R2**: peri-terrestrial | **C2**: indirect | **H1**: Type I |
| **L3**: intelligent | | **E3**: technological / post-biological | | **R3**: solar system | | **H2**: Type II |
| | | | | **R4**: solar neighbourhood | | **H3**: Type III |
| | | | | **R5**: Milky Way Galaxy | | |
| | | | | **R6**: extra-galactic | | |

**Figure 2:** The LSEARCH Sector/Factor array; a minimalist morphology for contact scenarios

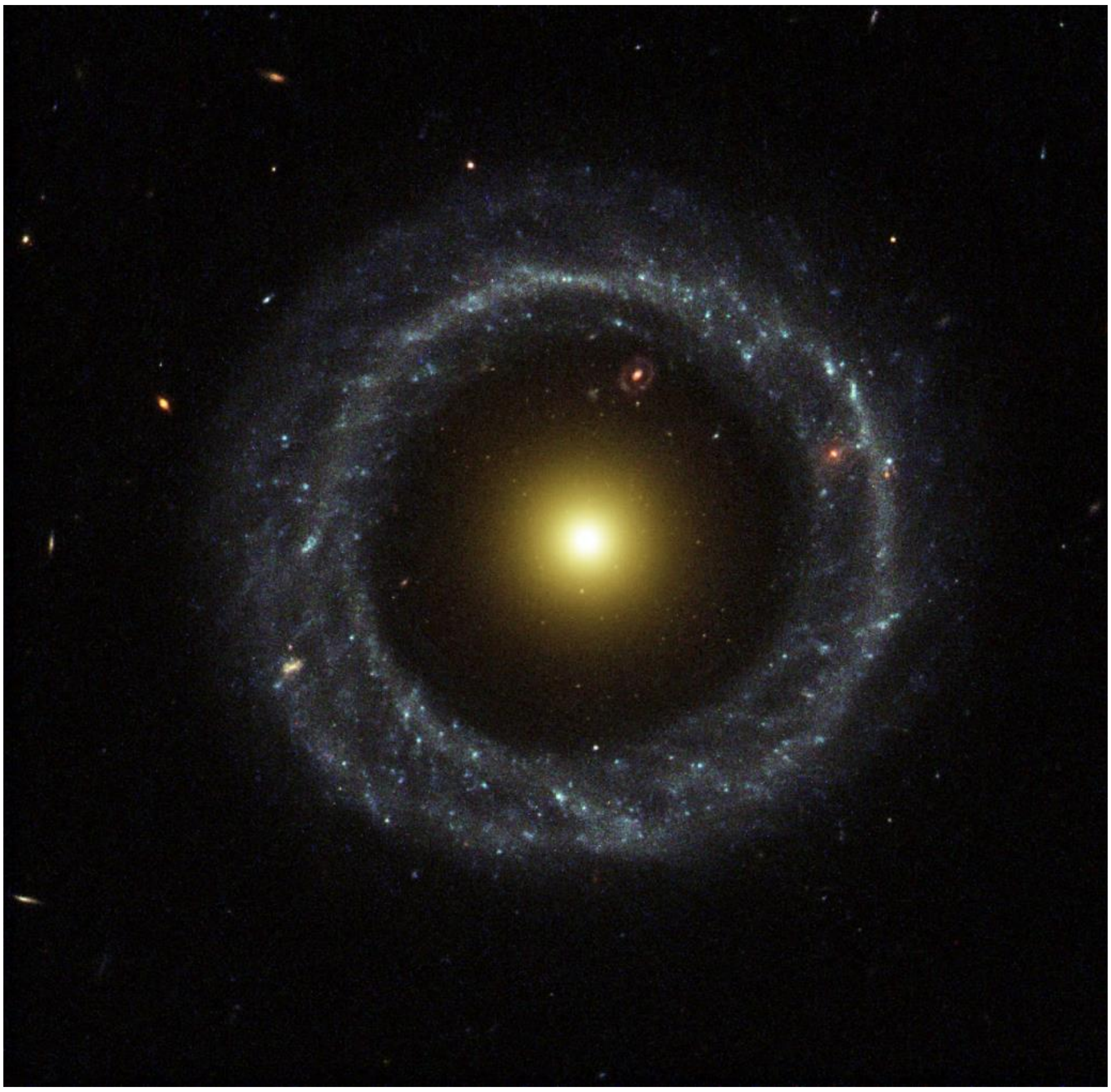

**Figure 3:** Hoag's Object (PGC54559).
Image credit: NASA & The Hubble Heritage Team (STScI/AURA). Acknowledgement: Ray A. Lucas (STScI/AURA).

---

## About the author:

**Joseph Voros** is a Senior Lecturer in Strategic Foresight at Swinburne University of Technology in Melbourne, Australia. He has a PhD in theoretical physics, in which he worked on mathematical extensions to the General Theory of Relativity. Following several years in Internet-related companies (including a stint at the legendary Netscape Communications in Silicon Valley) he became a professional futurist. His teaching and research interests are broadly multi-disciplinary and include: theories and models of social change; the long-term past and future of humankind; Astrobiology/SETI; and the broad sweep of cosmic evolutionary history. Three of his research articles have won excellence awards, including an Outstanding Paper Award in 2010 for 'Morphological Prospection: Profiling the Shapes of Things to Come' (2009). He is a member of the World Futures Studies Federation, an academic member of the Big History Institute, a full member of the Association of Professional Futurists, and a current Board member of the International Big History Association. He also serves on the editorial boards of the journal *Foresight*; the *European Journal of Futures Research* and the *Journal of Big History*.